\documentclass[aps,pra,twocolumn,10pt,letterpaper,superscriptaddress]{revtex4-2}

\usepackage{blindtext}
\usepackage{natbib}
\usepackage[T1]{fontenc}
\usepackage[english]{babel}
\usepackage{amsmath}
\usepackage{amsfonts}
\usepackage{amssymb}
\usepackage[colorlinks=true,linkcolor=blue,urlcolor=blue,citecolor=blue]{hyperref}
\usepackage{lineno}
\usepackage{braket}
\usepackage{wrapfig}
\usepackage{todonotes}

\newcommand{\mi}{\mathrm{i}}
\newcommand{\alg}[1]{\mathfrak{#1}}
\newcommand{\op}[1]{\hat{#1}}
\newcommand{\comm}[3]{[#1,#2]=#3}

\begin{document}
\title{Integrated optical wave analyzer using the discrete fractional Fourier transform}
\author{A.~R. Urz\'ua}
\email[Corresponding author:\,]{alejandro@icf.unam.mx}
\affiliation{Instituto de Ciencias F\'isicas, Universidad Nacional Aut\'onoma de M\'exico, Apartado Postal 48-3, 62251 Cuernavaca, Morelos, Mexico}

\author{I. Ramos-Prieto}
\affiliation{Instituto Nacional de Astrof\'isica, \'Optica y Electr\'onica, Calle Luis Enrique Erro No. 1, Santa Mar\'ia Tonantzintla, Puebla, 72840, Mexico}

\author{H.~M. Moya-Cessa}
\affiliation{Instituto Nacional de Astrof\'isica, \'Optica y Electr\'onica, Calle Luis Enrique Erro No. 1, Santa Mar\'ia Tonantzintla, Puebla, 72840, Mexico}

\begin{abstract}
Within the expansive domain of optical sciences, achieving the precise characterization of light beams stands as a fundamental pursuit, pivotal for various applications, including telecommunications and imaging technologies. This study introduces an innovative methodology aimed at reconstructing the Wigner distribution function of optical signals; a crucial tool in comprehending the time-frequency behavior exhibited by these signals. The proposed approach integrates two robust mathematical tools: the discrete realization of fractional Fourier transform, and the propagator of the quantum harmonic oscillator in waveguide arrays. This integration offers a direct and efficient method for characterizing optical signals by reconstructing their Wigner distribution function in the scope of integrated optics. We provide evidence of how having knowledge of the signal propagation amid the phase-space reconstruction, has desirable advantages in respect to only knowing the signal state.
\end{abstract}
\date{\today}
\maketitle
\section{Introduction}
In the realm of physics, the Wigner distribution function~\cite{Wigner_1932} holds a crucial role in depicting quantum particle wave functions by utilizing two conjugate variables, such as position and momentum, or number and phase~\cite{Moya2003,Perez_2016,Ramos_2020}. Its significance transcends quantum mechanics, finding extensive applications in signal processing~\cite{Bastiaans_1978,Cohen_1989,Dragoman_2005} and classical optics~\cite{Bastiaans_1979,Alonso_2009,Alonso_2011}, owing to its close association with the Fourier transform.

Introduced in 1980 by Namias using an operator-based approach~\cite{Namias_1980}, the continuous fractional Fourier transform notably encompasses the conventional Fourier transform as a specific instance. Inheriting crucial properties from its classical counterpart, the fractional Fourier transform has revealed numerous applications across diverse domains, including signal analysis, pattern recognition, both paraxial and non-paraxial optics, and optical image encryption~\cite{Mendlovic_1993,Ozaktas_1993,ozaktas2001fractional,Almeida_1994,Alieva_1994,LOHMANN_1996,Tao_2008, Ramos_2023}. Studies from a tomographic perspective have demonstrated that applying the Radon transform to the Wigner distribution function defines the squared modulus of the fractional Fourier transform~\cite{Radon,Lohmann1994,Wood_1994,Mendlovic_1996optical}.

In optical models, achieving both the complete and fractional Fourier transforms involves employing sets of lenses~\cite{Unnikrishnan_2000,Wang_2007experimental}. In this setup, an input image undergoes transformation either at the focal point or at an intermediate position along the optical path. However, in discrete and finite scenarios, successful implementation of the fractional Fourier transform has been achieved using waveguide arrays, where the coupling between adjacent waveguides is governed by the angular momentum operator $\hat{J}_x$ \cite{Armando_jx}. Among various implementations of the discrete fractional Fourier transform \cite{kbna, Barker2000, Jafarov2011, Atakishiyev1997}, we specifically selected this operational approach to perform tomographic reconstructions of the Wigner distribution function. In this context, finite or semi-infinite waveguide arrays (integrated optics) have been situated within the realm of classical-quantum analogies, opening doors to pioneering investigations into phenomena governed by quantum theory~\cite{Christodoulides_2003,Longhi_2009,Armando_CorrelationsG,Aleahmad,Abouraddy,Ancheyta_2017,Ramos_2021}. Within this framework, our study unfolds by identifying the discrete harmonic oscillator in terms of the angular momentum operator $\hat{J}_x$ and, notably, recognizing its propagator as the discrete fractional Fourier transform.

However, as the field is discretely diffracted along the propagation distance until completely recovered after a complete period, a new question arises: Can we take this as a \emph{sinogram} (or a parallel projection) in the Radon sense? The sense is where rows are detector position numbers along a line, and columns are angles; so it is possible to perform filtered back projections to obtain a discrete version of the Wigner distribution function  of the input field. This is no other thing but using parallel projection tomographic techniques. So, the answer is yes, with some cautions on the scale factor side of the final representation of the WDF \cite{Lohmann1994}. We emphasize, albeit the calculation of a conjugate variables' joint phase-space can be done directly having knowledge of the initial wavefunction, there's an advantage in having their propagation in a related space, like waveguides, where we can preview some features that lead to characteristics inherited to the Wigner function.

The manuscript is organized as follows. In \S\ref{seccion1} we take an account in the construction of the discrete fractional Fourier transform as a discrete realization of the Discrete Quantum Harmonic Oscillator, using the $\hat{J}_{x}$ eigenfunctions, and their proper implementation in an evanescent waveguides arrays. Here, we state how the functions that model the propagation of light (and information) propagates along the angular parameter of the discrete transformation. In \S\ref{seccion2}, we implement the Wigner-Radon protocol using the backprojection on the angle-line space spawned by the (discrete fractional) propagation. Here we give some examples of reconstruction of the Wigner function for selected initial wavefunction. Thus, we can respond the question arose above: the discrete fractional Fourier transform can be seeing as a proper sinogram, enabling an analysis of optical profiles. In \S\ref{sec_3} we give our conclusions and perspectives on the method and protocol proposed in this work.

\section{Discrete fractional Fourier transform}\label{seccion1}
In this section, we explore the concept of the discrete fractional Fourier transform (DFrFT), which is founded upon a set of eigenvectors associated with the Discrete Quantum Harmonic Oscillator (DQHO), serving as the discrete counterpart to the well-established Hermite-Gauss functions \cite{suslov91,kbna,Candan}. It is of paramount importance to reiterate that the wavefunction of a Quantum Harmonic Oscillator (QHO) at a time $t$ is precisely described by the propagator, as outlined in Appendix \ref{A} \cite{Namias_1980,Dattoli_1998,Kutay_2002},
\begin{equation}\label{Fraccionaria_integral2}
\psi(x;t)=\int\limits_{-\infty}^{\infty}K(x,\mu;t) \tilde{\psi}(\mu;0)d\mu,
\end{equation}
where $\tilde{\psi}(\mu;0)$ is the FT of the initial condition, and
\begin{align}\label{Kernel}
K(x,\mu,t)=\frac{e^{-\frac{\mi}{2}\tan(t)x^2}}{\sqrt{\cos(t)}}e^{2\mi\pi^2\tan(t)\mu^2}e^{-\mi\frac{2\pi}{\cos(t)}\mu x},
\end{align}
is the transformation kernel, known as the Green's function of the QHO. Two critical aspects require emphasis: a) Despite the initial condition being set at $t=0$, the boundary conditions on $\mu$ span all possible points within the configuration space. This leads to the FT of $\tilde{\psi}(\mu;0)$, which essentially represents the initial wave function.
b) When time reaches $t=\pi/2$, the kernel of the previous transformation simplifies into the FT. This holds great importance, as it yields fractional orders of the FT within the interval from $0$ to $\pi/2$.

\begin{figure}
	\includegraphics[width=\linewidth]{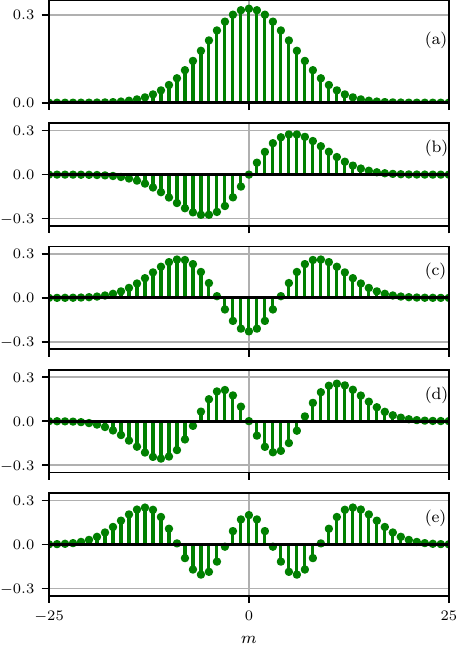}
	\caption{(a)-(e) The first five eigenvectors of the DQHO are depicted, respectively. The spatial center of these eigenvectors, denoted by $\psi_n^m(q)$ (with $-j\leq m\leq j$), is governed by the parameter $q$. Specifically, in this illustration, we set $q=0$, and the system exhibits an odd dimensionality with $N=2j+1$ (with $j=25$).}\label{Fig_1}
\end{figure}

Within this perspective, it is noteworthy that previous research \cite{Armando_jx} has shown that a feasible discrete and finite version of the QHO can be derived by considering generators associated with the $\alg{su}(2)$ rotation algebra within a setup of evanescent coupled waveguide arrays, thus defining an odd-dimensional Hilbert space \cite{Orzag, Dattoli_1998}. Vectors in this space, denoted as $\ket{j,m}$, possess finite support. Specifically, for given values of $j$ and $m$, these vectors have a spectral range extending from $-j$ to $j$, resulting in a space dimensionality of $N = 2j + 1$. In this context, the generators follow the standard commutation relations, i.e., $\comm{\op{J}_{k}}{\op{J}_{l}}{\mi\epsilon_{klm}\op{J}_{m}}$. Thus, we propose the utilization of the following Hamiltonian,
\begin{equation}\label{Hamiltonian_DQHO}
\hat{H} = \hat{J}_x.
\end{equation}
This choice governs not only the system's entire dynamics but also faithfully replicates the essential characteristics of the QHO. Furthermore, it naturally gives rise to fractional orders of the FT within discrete and finite spaces~\cite{Armando_jx}. Calculating the matrix elements of this Hamiltonian in the diagonal basis of $\op{J}_{z}$, they can be expressed as follows:
\begin{equation}\label{J_x_elements}
\begin{split}
\bra{j,n}\hat{J}_x\ket{j,m}=&\frac{\kappa_0}{2}\bigg[\sqrt{j(j+1)-m(m+1)}\delta_{n,m+1}\\
&+\sqrt{j(j+1)-m(m-1)}\delta_{n,m-1}\bigg],
\end{split}
\end{equation}
where $\kappa_0$ is a scaling factor. Therefore, we are dealing with odd-dimensional matrices $N\times N$ that have off-diagonal elements.
In this context, direct diagonalization methods can be employed to obtain the appropriate eigenvector basis of Eq.~\eqref{Hamiltonian_DQHO} \cite{Orzag,Armando_jx}. Analogously to how Hermite-Gauss polynomials serve as eigenfunctions of the QHO, we can construct eigenvectors for this discrete and finite oscillator using functions that support the requisite indexing. Consequently, the $m$-th component of the resulting eigenfunctions can be expressed as follows:
\begin{equation}\label{eigenstates}
\begin{split}
\psi_n^m(q) = & 2^n\sqrt{\frac{(j+n)!(j-n)!}{(j+m-q)!(j-(m-q))!}}\\
&\times P_{j+n}^{m-q-n,-m+q-n}(0),
\end{split}
\end{equation}
where $P_k^{\alpha,\beta}(x)$ represents the Jacobi polynomials of order $k$ \cite{Orzag}. In Fig.~\ref{Fig_1}, the first five eigenvectors of the DQHO, Eq.~\eqref{eigenstates}, are plotted, illustrating their discrete approximation to the well-known Hermite-Gauss polynomials of the QHO. It is worth noticing that there exists a sort of realizations for the DFrFT, whose count those constructed by 1) sampling the harmonic oscillator eigenfunctions, the ``Taipei basi'' \cite{Pei1997, SooChangPei1999}; 2) sums of periodically displaced oscillator eigenfunctions, the ``Metha basis'' \cite{Mehta1987}; 3) the vibrating chain lattice model, the ``the Ankara lattice'' \cite{Candan2000, Barker2000}; 4) and the $\mathfrak{su}$(2) model of the discrete oscillator, the ``Wolf-Fourier-Kravchuk'' realization. Despite, every one of these versions of the DFrFT attains some peculiar properties, the use of the $\hat{J}_{x}$ realization is chosen as the most closed related to the propagation in waveguides arrays.

\begin{figure}
	\includegraphics[width=\linewidth]{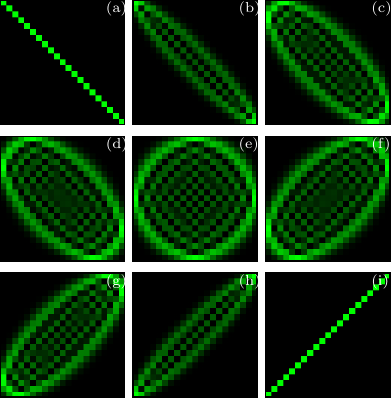}
	\caption{Various fractional orders for the DFrFT kernel, defined by Eq.~\eqref{Fractional_Operator}, are showcased. These orders are associated with specific values of $Z$, including $0,\frac{\pi}{8},\frac{\pi}{4},\frac{\pi}{3},\frac{\pi}{2},\cdots,\pi$, respectively. (e) Notably, at $Z=\pi/2$ in this sequence represents the kernel of the DFT.}\label{Kernel_fig}
\end{figure}

With Eq.~\eqref{Fraccionaria_integral2} and Eq.~\eqref{Kernel} (also refer to Appendix \ref{A}), we now direct our attention to the dynamic evolution of the DQHO. Naturally, this is governed by a Schr\"odinger-like equation,
\begin{equation}\label{Schrodinger_equation}
i\frac{\partial}{\partial Z}\ket{\psi(Z)}=\hat{J}_x\ket{\psi(Z)}.
\end{equation}
In this context, $\ket{\psi(Z)}=\sum_{m=-j}^jE_m(Z)\ket{j,m}$ denotes a vector that is defined in a Hilbert space with dimensions $N = 2j+1$. This serves as a spectral decomposition of the general functions $\ket{\psi}$ in relation to the basis $\ket{j,m}$, and their associated amplitudes $E_{m}(Z)$. Additionally, we have made a change, without loss of generality, substituting the time variable $t$ with a propagation variable $Z$. Hence, the remaining challenge lies in devising an efficient method to apply the propagation operator to the initial state vector at $Z=0$,
\begin{equation}\label{Solution_S}
\ket{\psi(Z)}=e^{-\mi Z\hat{J}_x}\ket{\psi(0)}.
\end{equation}
Although the propagation operator is given in terms of the angular momentum operator $\hat{J}_x$, it is challenging to find a specific representation, since it is a tridiagonal finite matrix given by Eq.~\eqref{J_x_elements}. In analogy to the Green's function approach for the QHO Eq.~\eqref{Kernel}, the evolution operator of the DQHO is the DFrFT. Furthermore, a similarity transformation employing the eigenfunctions, as described in Eq.~\eqref{eigenstates}, enables the determination of the Green's function for the DQHO~\cite{Armando_jx,Orzag},
\begin{equation}
\begin{split}\label{Kernel_E}
\mathcal{K}_{n,m}(Z)&=\mi^{n-m}\sqrt{\frac{(j+n)!(j-n)!}{(j+m)!(j-m)!}}[\sin(Z/2)]^{m-n}\\
&\times[\cos(Z/2)]^{-m-n}P_{j+n}^{m-n,-m-n}(\cos(Z)).
\end{split}
\end{equation}
Similarly to the continuous scenario, the evolution operator coincides with the DFrFT, as expressed by
\begin{equation}\label{Fractional_Operator}
\hat{\mathcal{F}}_Z=\sum_{m,n=-j}^{j}\mathcal{K}_{n,m}(Z)\ket{j,n}\bra{j,m}.
\end{equation}
Fig.~\ref{Kernel_fig} illustrates the square modulus of the DFrFT for various fractional orders determined by the variable $Z$. For each fractional order ranging from $0$ to $\pi$, this provides a transformation of the finite and discrete object within the context of a converging lens. This comprehensive analysis enables a more profound understanding of the optical properties of the system across this range of fractional orders.

\begin{figure}
	\includegraphics[width=\linewidth]{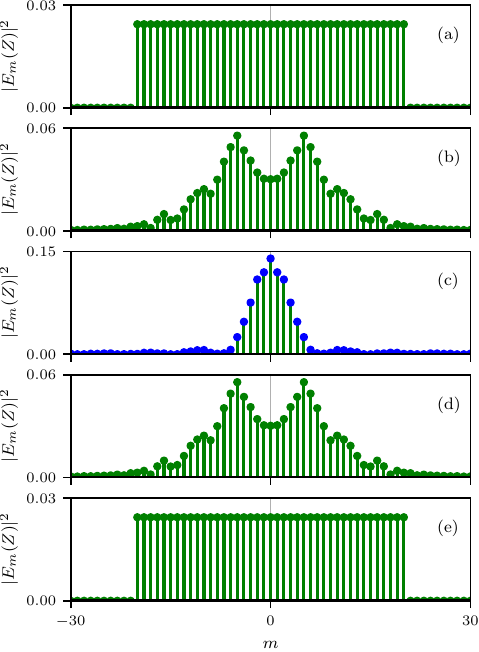}
	\caption{(a) Initial condition that represents the sampling of a normalized rectangular aperture (for a unit evolution). (b) and (c) represent the evolution of the initial condition at distances $Z=\pi/5$ and $Z=\pi/3$, respectively. Figure (d) with $Z=\pi/2$, is the square modulus of the discrete Fourier Transform (DFT) of a rectangular function.}\label{rec_fig}
\end{figure}

To exemplify the DFrFT operator $\hat{\mathcal{F}}_Z$, we consider a representative scenario where the initial condition is determined by the sampling of a normalized rectangular function as specified in Eq.~\eqref{Solution_S}. Subsequently, we generate plots depicting the squared modulus for varying propagation distances. In Fig.~\ref{rec_fig}, it is evident that each discrete value indexed by $m$ in the DFrFT at distance $Z$ is calculated using the following equation:
\begin{equation}
E_m(Z) = \braket{j,m|\hat{\mathcal{F}}_Z|\psi(0)}.
\end{equation}
Where $\ket{\psi(0)}=\sum_{k=-j}^{j}E_k(0)\ket{j,k}$ signifies the initial condition. It is worth emphasizing, as previously mentioned, that at $Z=\pi/2$, the kernel of the transformation simplifies to the DFT, as showcased in Fig.~\ref{rec_fig}~(d). In the scenario of a one-dimensional rectangular aperture, this leads to a discrete $sinc(x)$ function.

In closing this section, it is imperative to underscore that our ability to obtain the DFrFT extends beyond arbitrary functions; we can also apply it to quantum states. This capability arises from the inherent connection between the evolution operator and the QHO. Theoretically, there are no limitations on the types of initial states we can employ, as long as they meet the requirements of proper sampling and satisfy the Nyquist criterion~\cite{Nyquist}, ensuring the well-defined nature of the DFrFT.

\section{The Wigner-Radon transform in photonic lattices}\label{seccion2}
The intricacy associated with Wigner function reconstruction via the inverse Radon transform (IRT) primarily stems from the task of obtaining parallel projections~\cite{Lohmann1994,Wood_1994,Tao_2008}. This demanding process necessitates precise alignment and collection of data from various angles, making it a pivotal aspect in the successful retrieval of the WDF. In this section, we use the Radon transform scheme, introduced by Johann Radon \cite{Radon}, where the parallel projections of an object or function of two variables indirectly provide its internal structure. In the specific case of parallel projections of the WDF (function of two variables), these turn out to be the square modulus of the FrFT \cite{Lohmann1994}. Within a discrete and finite framework, the implementation of the DFrFT in photonic arrays hinges on the propagation distance, which dictates the fractional order of the transform~\cite{Armando_jx}. This relationship mirrors the connection observed in the continuous domain between the DQHO and DFrFT.

\begin{figure}
	\includegraphics[width=\linewidth]{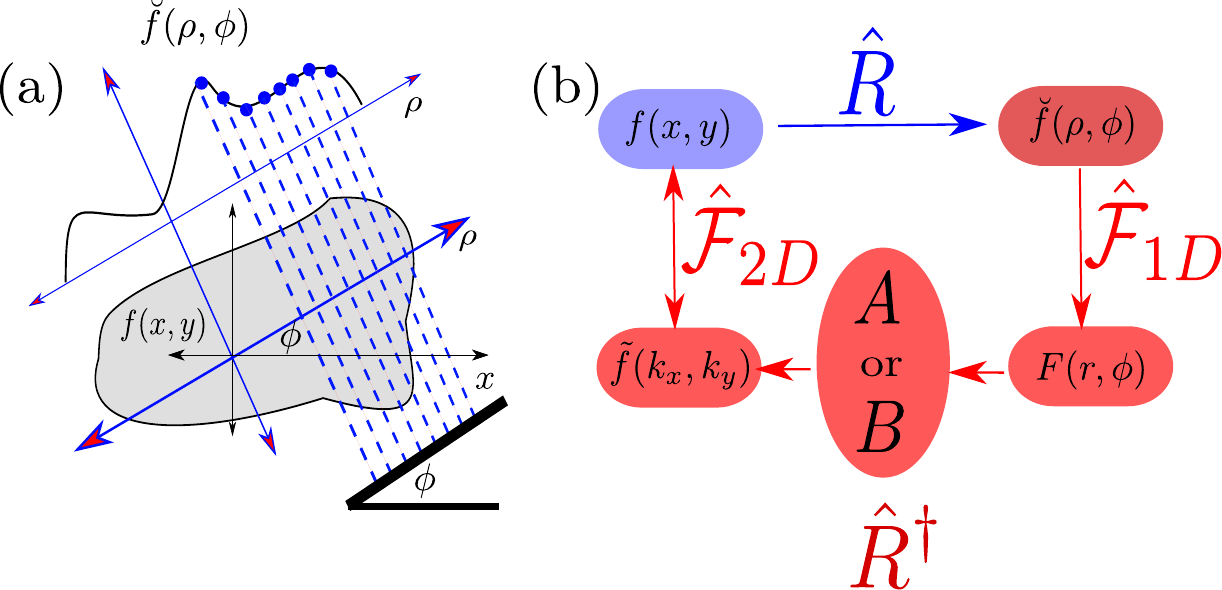}
	\caption{
(a) Schematic illustration depicting the RT of a two-variable function $\breve{f}(\rho,\phi)=\hat{R}\{f(x,y)\}$. Each blue point symbolizes the outcome of integration along the dashed line defined by the Dirac delta function. (b) Once all the parallel projections $\breve{f}(\rho,\phi)$ are acquired, two highly valuable methods exist for reconstructing the function $f(x, y)$: (A) The Fourier slice theorem, and (B) a method that incorporates filtering within the same framework. Both techniques initially employ the one-dimensional transform and culminate in the two-dimensional Fourier transform}\label{Radon_fig}.
\end{figure}

The RT is an integral transform that arises from the parallel projection of a two-variable function $f(x,y)$. It is mathematically defined through the following line integral,
\begin{equation}
\begin{split}\label{Radon_Definiton}
\breve{f}(\rho, \phi) &= \hat{R}\{f(x, y)\} \\
&= \int\limits_{-\infty}^\infty \int\limits_{-\infty}^\infty f(x, y) \delta(\rho - x\cos\phi - y\sin\phi) dx dy,
\end{split}
\end{equation}
where $\hat{R}$ denotes the action of the RT on the function $f$, $\rho$ represents the abscissa of the rotated system, while $\phi$ denotes the rotation angle (refer to Fig.~\ref{Radon_fig}~(a)). Consequently, the number of RT instances corresponds to the number of applied rotation angles. 
In this regard, Lohmann and Soffer have provided evidence that the FrFT indeed arises from the parallel projections of the WDF~\cite{Lohmann1994}. Therefore, applying the RT to the WDF yields the following,
\begin{equation}\label{Radon_Wigner}
\hat{R}\{W(x,y)\}=\int\limits_{-\infty}^{\infty}W(x,y)\delta(\rho-x\cos\phi-y\sin\phi)dxdy,
\end{equation}
where
\begin{equation}\label{Wigner}
W(x,y)=\int\limits_{-\infty}^{\infty}f(x+x'/2)f^*(x-x'/2)e^{-i2\pi x'y}dx',
\end{equation}
represents the integral version of the WDF for a given function $f(x)$. After performing a suitable change in the integration variables, employing Eq.~\eqref{Radon_Wigner} and Eq.~\eqref{Wigner}, Eq.~\eqref{Radon_Definiton} can be expressed in the following manner,
\begin{equation}\label{CFrFT}
\begin{split}
&\breve{f}(\rho,\phi)=\hat{R}\left\{W(x,y)\right\}\\&=\bigg|\int\limits_{-\infty}^\infty\frac{e^{-\frac{\mi}{2}\tan(\phi)\rho^2}}{\sqrt{\cos(\phi)}}e^{2\mi\pi^2\tan(\phi)x^2}e^{-\mi\frac{2\pi}{\cos(\phi)}x \rho} f(x)dx \bigg|^2.
\end{split}
\end{equation}
From Eq.~\eqref{Kernel}, we can observe that the previous result can be recognized as the square modulus of the FrFT. This result has been previously demonstrated in~\cite{Lohmann1994}. As mentioned earlier, at $\phi=\pi/2$, the kernel of the transformation corresponds to that of the FT. Furthermore, the projection angle determines each fractional order. Eq.~\eqref{CFrFT} serves as the preamble to our discrete tool for reconstructing the WDF in photonic systems, enabling the characterization and analysis of any discretized signal.

Let us now explore the inverse problem, which is tackled by applying the IRT to the square modulus of the FrFT. Two primary approaches are commonly employed to derive the inverse transformation, both rooted in the Fourier Slice Theorem \cite{Bracewell697,Bruyant01102002}, also known as the Central Slice Theorem. In general, the reconstruction of the WDF or any other function requires parallel projections for each angle. As indicated by Eq.~\eqref{Radon_Wigner}, when considering one of the possible fractional orders, denoted as $\phi$, of the FT applied to a function $f(x)$, it becomes possible to derive the following:
\begin{equation}\label{WignerIRadon}
W(x,y)=\hat{R}^{-1}\{|\hat{\mathcal{F}}_{\phi}\{f(x)\}|^2\}.
\end{equation}
In this context, $\hat{R}^{-1}$ is used to represent the IRT operator, with $\phi$ lying within the range of $[0, \pi]$. Each fractional order corresponds to specific points in the phase space of the WDF. In this regard, the objective of the IRT is to recover or indirectly reconstruct information from a given signal. However, the computational implementation varies depending on the specific objective and is accomplished through various methods \cite{avinash1988principles,deans2007radon}: series and orthogonal functions, iterative approaches, direct Fourier methods, as well as signal space convolution and frequency space filtering methods. One of the most straightforward Fourier methods is based on the central slice theorem \cite{mersereau1974digital}, which establishes a relationship between two-dimensional and one-dimensional Fourier transforms through the RT (as depicted in Fig.~\ref{Radon_fig}(b)). Within this framework, it is possible to derive the discrete Wigner distribution (DWDF), either by utilizing the Fourier slice theorem or through the application of filtered back-projection, resulting in
\begin{equation}\label{R_DWDF}
\begin{split}
W(x_i,y_i)&=\frac{1}{\sqrt{(2N+1)(2M+1)}}\\
&\times\sum_{n=0}^{2N}\sum_{m=0}^{2M}F_n(Z_m)e^{2\mi\pi(\frac{nx_i}{2N}+\frac{my_i}{2M})},
\end{split}
\end{equation}
that represents the two-dimensional DFT of the discrete signal $F_n(Z)$, where
\begin{equation}\label{F_n}
F_n(Z_m)=\sum_{k=0}^{2N}I_k(Z_m)e^{-\mi\frac{2\pi}{N}nk}.
\end{equation}
Where, $x_i$ and $y_i$ denote points within the phase space, while $Z_m$ assumes the role of $\phi_m$, where each value of $m$ corresponds to a distinct fractional order. Furthermore, $I_k(Z_m)$ signifies the squared modulus of the $k$-th point within the $m$-th fractional order during the sampling process of a function $f(x)$. Here, $N$ and $M$ are determined by the number of sampling points in the input signal and the number of parallel projections, respectively. 

\begin{figure}
	\includegraphics[width=\linewidth]{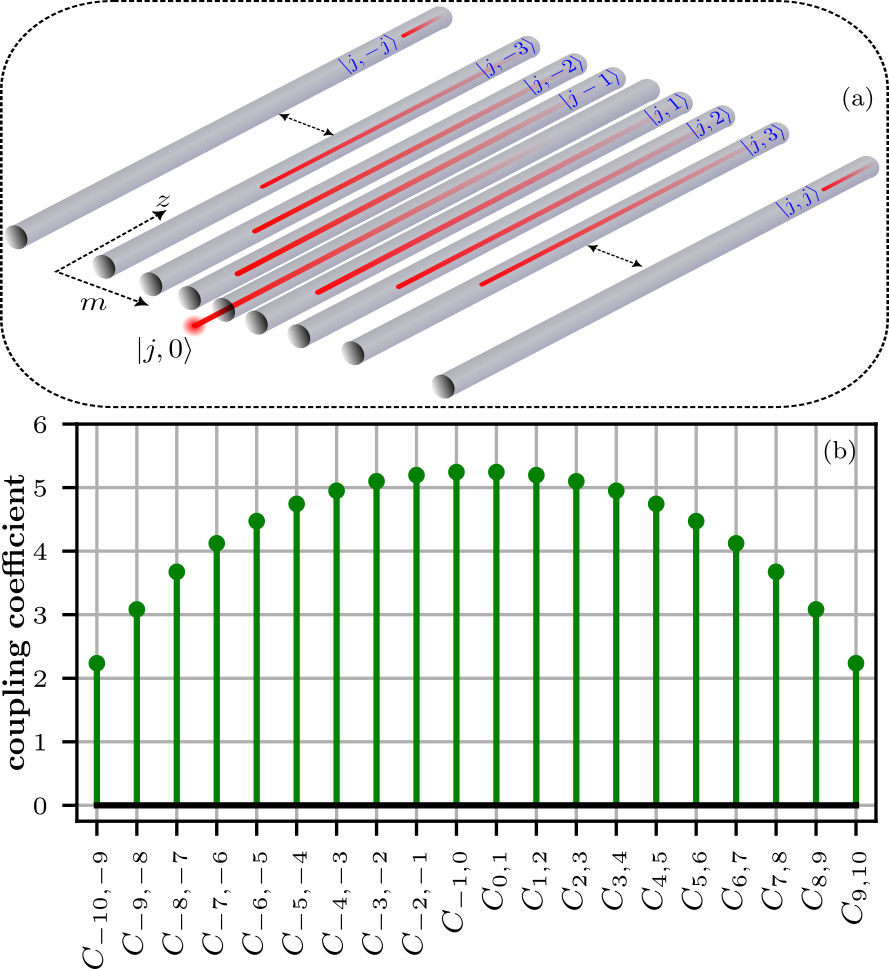}
	\caption{(a) Each guide labels an element of the base of the finite Hilbert space of the DQHO, where $j$ is an integer, and $-j\leq m\leq j$ is the waveguide number. (b) Coupling distribution obtained from the matrix elements $\braket{j,n|\hat{J}_x|j,m}$, in an array of $N=2j+1$ waveguides, with $j=15$.}\label{H_fig}
\end{figure}

Now, within the discrete and finite framework defined by the DQHO, and considering the scenario of waveguide arrays as governed by the Hamiltonian Eq.~\eqref{Hamiltonian_DQHO}, we have identified an opportunity for WDF reconstruction. As previously demonstrated, the propagation operator Eq.~\eqref{Fractional_Operator} corresponds to the DFrFT, implying distinct fractional orders for each propagation distance. In the context of the tight-binding limit for a group of $N = 2j + 1$ waveguides with coupling coefficients defined by $\hat{J}_x$ (as shown in Fig.~\ref{H_fig}), the light intensity in each waveguide at a specific propagation distance, $Z$ (in fractional order), can be expressed by considering a generic initial condition $\ket{\psi(0)}$,
\begin{equation}\label{E_FractionalE}
I_k(Z)=|\braket{j,k|\psi(Z)}|^2=|\braket{j,k|\hat{\mathcal{F}}_Z|\psi(0)}|^2.
\end{equation}
Here $\ket{\psi(0)}$ can be regarded as the sampling of some function $f(x)$ decomposed in the discrete basis, $\ket{\psi(0)} = \sum_{m=-j}^{j}E_m(0)\ket{m}$, where $E_m(0)$ represents the amplitudes of the generic sampled signal. Consequently, the light intensity in the $k$-th waveguide undergoes evolution over a distance $Z$, and this evolution is determined by the DFrFT operator $\hat{\mathcal{F}}_Z$, as described in Eq.~\eqref{Fractional_Operator}. Given that the propagation of light within a waveguide array featuring parabolic-type coupling \cite{Armando_jx,Armando_multiphotonjx} is governed by the Schrödinger equation, Eq.~\eqref{Schrodinger_equation}, we can express the evolution of intensity in the array-$\hat{J}_x$ as a matrix:
\begin{equation}\label{matrix_R}
\begin{pmatrix}
I_{-j}(Z_0)&I_{-j}(Z_1)&\cdots&I_{-j}(Z_{n-1})&I_{-j}(Z_n)\\
I_{-j+1}(Z_0)&I_{-j+1}(Z_1)&\cdots&I_{-j+1}(Z_{n-1})&I_{-j+1}(Z_n)\\
\vdots&\vdots&\vdots&\vdots&\vdots\\
I_{0}(Z_0)&I_{0}(Z_1)&\cdots&I_{0}(Z_{n-1})&I_{0}(Z_n)\\
\vdots&\vdots&\vdots&\vdots&\vdots\\
I_{j-1}(Z_0)&I_{j-1}(Z_1)&\cdots&I_{j-1}(Z_{n-1})&I_{j-1}(Z_n)\\
I_{j}(Z_0)&I_{j}(Z_1)&\cdots&I_{j}(Z_{n-1}&I_{j}(Z_n)
\end{pmatrix}.
\end{equation}
To generate each column, a dedicated waveguide array is constructed with the desired propagation distance $Z_n$, and the output intensities in all waveguides are measured. An alternative approach involves doping the waveguides with a suitable fluorescent material and watching the fluorescence intensity along the propagation, with the observation taking place from above the waveguide chip~\cite{Szameit:07}; Fig.~\ref{H_fig}~(a) represents schematically a finite set of identical evanescent coupled waveguides, where $j$ is an integer that determines the dimension of the Hilbert space or the total number of waveguides $N=2j+1$, while Fig.~\ref{H_fig}~(b) shows the matrix elements of the angular momentum operator $\hat{J}_x$, Eq.~\eqref{J_x_elements}, that follow a growth of  parabolic type. Under these considerations, we are in the possibility of using the DFrFT in a waveguide beam splitter scenario.

\begin{figure}
	\includegraphics[width=\linewidth]{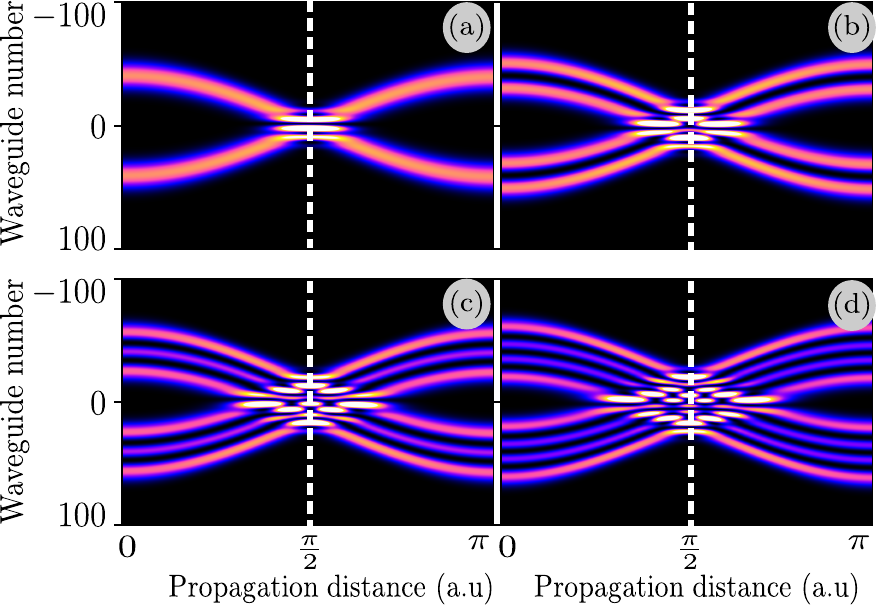}
	\caption{We plot the propagation of light in the array $\hat{J}_x$ for the superposition of eigenstates of the DQHO defined by equation \eqref{eigenstates}. (a) $\psi_{0}^m(50)+\psi_{0}^m(-50)$. (b) $\psi_{1}^m(50)+\psi_{1}^m(-50)$. (c) $\psi_{2}^m(50)+\psi_{2}^m(-50)$. (d) $\psi_{3}^m(50)+\psi_{3}^m(-50)$. With $-50\leq m\leq 50$, and normalized, respectively.}\label{cats}
\end{figure}

To illustrate the WDF reconstruction process, we will explore four specific scenarios involving the superposition of eigenstates as defined by Eq.~\eqref{eigenstates}, which encompass distinct $q$ values. Fig.~\ref{cats} showcases linear combinations or superpositions of states, commonly regarded as Schr\"odinger cat states \cite{Schrodinger_1935,GSCatS}. The fractional order in each guide is determined by Eq.~\eqref{E_FractionalE}, which, at $Z = \pi / 2$, transforms into the discrete integer representation of the initial condition. This transformation, known within the framework of parallel projections, forms the \emph{sinogram} depicting the studied signal. In the context of the Radon-Wigner transform, the \emph{sinogram} is defined through the square modulus of the various orders of the fractional discrete Fourier transform (representing the columns in Eq.~\eqref{matrix_R}). These orders are contingent upon the propagation distance $Z$. Once all the fractional orders ($Z\in[0,\pi]$) are obtained, it becomes feasible, employing either the Fourier slice theorem or filtered back-projection techniques, to derive the discrete Wigner distribution. This derivation involves Eqs.~\eqref{R_DWDF} and \eqref{F_n}. Notably, it's imperative to recognize that each $I_k(Z_m)$ signifies the intensity in the $m$-the waveguide as a function of the propagation distance $Z_m$~(refer to appendix \ref{C}).

\begin{figure}
	\includegraphics[width=\linewidth]{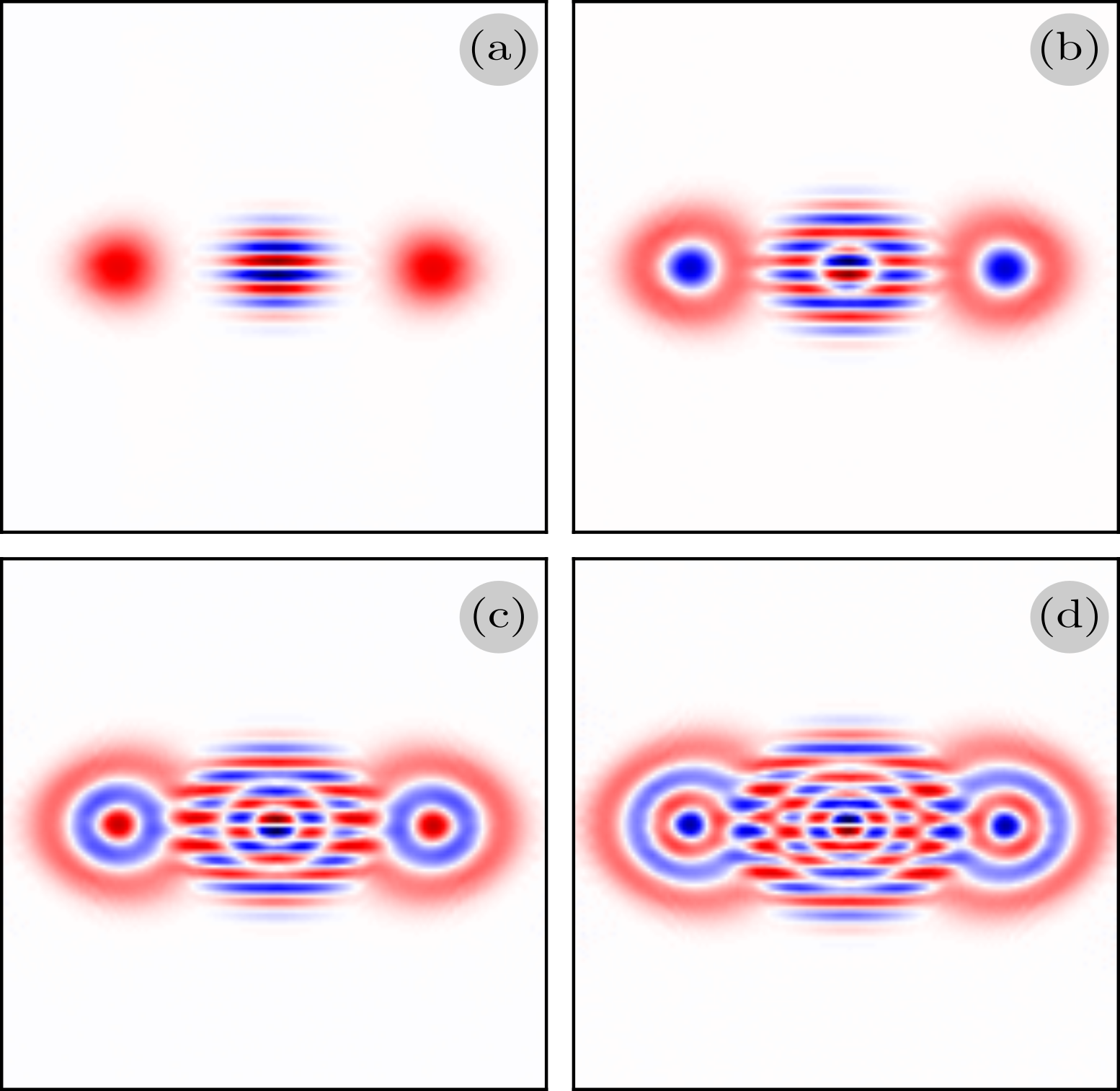}
	\caption{Reconstruction of the WDF for the following initial conditions: (a) $\psi_{0}^m(50)+\psi_{0}^m(-50)$. (b) $\psi_{1}^m(50)+\psi_{1}^m(-50)$. (c) $\psi_{2}^m(50)+\psi_{2}^m(-50)$. (d) $\psi_{3}^m(50)+\psi_{3}^m(-50)$.}\label{RadonWigner}
\end{figure}

Finally, using Eq.~\eqref{R_DWDF} in each of the cases shown in Fig.~\ref{cats} it is possible to obtain the reconstruction of the discrete Wigner distribution function. In Fig.~\ref{RadonWigner}, by using the scikit-image toolbox \cite{scikit-image}, we have reconstructed the WDF associated with the superposition of eigenstates from QDHO, so-called generalized Schr\"odinger cat states \cite{GSCatS}. 

\section{Conclusions}\label{sec_3}
We have demonstrated that, akin to the continuous case of the quantum harmonic oscillator's propagator identified as the Fractional Fourier Transform (FrFT), its discrete version accommodates fractional orders of the Discrete Fourier Transform (DFT). This discrete representation, integrated within an array of waveguides marked by lines and angles, allows us to reinterpret the progression of discretized input states as parallel projections conforming to the Radon transform's framework. Consequently, this reinterpretation facilitates its application within integrated circuits. It's noteworthy that the discretization of the quantum harmonic oscillator lacks a unique definition, leading to several approaches to obtaining appropriate discrete eigenfunctions that mirror those in the continuous scenario. Nevertheless, each one of these realizations has its scope and applicability. Selecting one, as we do in this work, can respond to the specifics of the system.

The Wigner function, pivotal in signal theory and quantum mechanics, unveils the concurrent time-frequency evolution within a signal. Despite its computationally intricate nature and the challenge in direct interpretation, the Fractional Fourier Transform and the Radon inverse emerge as compelling tools. They facilitate an efficient reconstruction of the Wigner function from data in both the frequency domain and spatial realms, proving pivotal in signal processing and quantum optics. And, as stated at the beginning, and demonstrated with examples, there's a clear advantage, quantitative and qualitative, to have the propagation of an initial wavefunction, where we can discern features that later will be inherited to the Wigner phase-space.

Moreover, we illustrated the quantum harmonic oscillator's propagator, conceptualized as a Fractional Fourier Transform (FrFT), finding practical utility in waveguide arrays, such as arrangements resembling $\hat{J}_x$. This realization, allows us to reinterpret the evolution of discretized states as projections within the Radon transform framework, simplifying their integration into integrated circuits. It's notable that discretizing the harmonic oscillator offers multiple approaches to derive discrete functions mimicking its continuous behavior. We hope these findings turn out to be useful for the community who wants to understand the connection between the concepts presented.

\appendix
\section{\label{A} Continuous fractional Fourier transform }
The Schr\"odinger equation of the quantum harmonic oscillator with $m = 1$, $k = 1$, and $\hbar = 1$, is
\begin{equation}
i\frac{\partial}{\partial t}\psi(x;t)=\frac{1}{2}(\hat{p}^2+\hat{x}^2)\psi(x;t).
\end{equation}
and therefore, the wave function at time t is 
\begin{equation}
\psi(x;t)=e^{-\mi\frac{t}{2}(\hat{p}^2+\hat{x}^2)}\psi(x;0),
\end{equation}
and using the fact that \cite{Dattoli_1998}
\begin{equation}
e^{-\mi\frac{t}{2}(\hat{p}^2+\hat{x}^2)}=e^{-\mi\frac{t}{2}\tan(t)\hat{x}^2}e^{-\frac{\mi}{2}\ln[\cos(t)](\hat{x}\hat{p}+\hat{p}\hat{x})}e^{-\frac{\mi}{2}\tan(t)\hat{p}^2},
\end{equation}
in addition to the following relationships
\begin{align}
\begin{split}
e^{-\frac{\mi}{2}\tan(t)\hat{p}^2}\hat{x}e^{\frac{\mi}{2}\tan(t)\hat{p}^2}&=\hat{x}-\tan(t)\hat{p},\\
e^{-\frac{\mi}{2}\ln[\cos(t)](\hat{x}\hat{p}+\hat{p}\hat{x})}\hat{x}e^{+\frac{\mi}{2}\ln[\cos(t)](\hat{x}\hat{p}+\hat{p}\hat{x})}&=\frac{\hat{x}}{\cos(t)},\\
e^{-\frac{\mi}{2}\ln[\cos(t)](\hat{x}\hat{p}+\hat{p}\hat{x})}\hat{p}e^{+\frac{\mi}{2}\ln[\cos(t)](\hat{x}\hat{p}+\hat{p}\hat{x})}&=\hat{p}\cos(t),\\
e^{-\frac{\mi}{2}\ln[\cos(t)](\hat{x}\hat{p}+\hat{p}\hat{x})}\hat{1}&=\frac{1}{\sqrt{\cos(t)}},
\end{split}
\end{align}
it is possible to obtain that the initial condition evolves as
\begin{equation}
\psi(x;t)=\frac{e^{-\frac{\mi}{2}\tan(t)\hat{x}^2}}{\sqrt{\cos(t)}}\psi\bigg(\frac{\hat{x}}{\cos(t)}-\sin(t)\hat{p};0\bigg).
\end{equation}
that when considering the inverse Fourier transform 
\begin{equation}
f(x)=\int\limits_{-\infty}^\infty d\mu \tilde{f}(\mu)e^{2\mi\pi\mu x},
\end{equation}
applied in the initial condition, it is obtained that
\begin{align}\label{Fraccionaria_integral}
\begin{split}
\psi(x;t)=&\frac{e^{-\frac{\mi}{2}\tan(t)x^2}}{\sqrt{\cos(t)}}\\
&\times\int\limits_{-\infty}^{\infty}d\mu\tilde{\psi}(\mu;0)e^{2\mi\pi^2\tan(t)\mu^2}e^{-\mi\frac{2\pi}{\cos(t)}\mu x}.
\end{split}
\end{align}
The last relation is known as the continuous fractional Fourier transform.

\section{\label{C}The Fourier slice theorem and filtered backprojection}
There are two representative numerical methods based on the Fourier slice theorem (Fourier central theorem) and the filtered back-projection to obtain the inverse Radon transform \cite{Bracewell697,Bruyant01102002}. To accomplish this, let's define the 2D Fourier transform of $f(x,y)$ as,
\begin{equation}
F(k_x,k_y)=\int\limits_{-\infty}^\infty f(x,y)e^{-2\mi\pi(k_xx+k_yy)}dxdy,
\end{equation}
while the inverse transform is given by
\begin{equation}\label{C2}
f(x,y)=\int\limits_{-\infty}^{\infty}F(k_x,k_y)e^{2\mi\pi(k_xx+k_yy)}dk_xdk_y.
\end{equation}
Therefore, considering polar coordinates
\begin{equation}
\begin{pmatrix}
k_x\\k_y
\end{pmatrix}=r\begin{pmatrix}
\cos\phi\\\sin\phi
\end{pmatrix},
\end{equation}
it is possible to obtain that
\begin{widetext}
\begin{align}
\begin{split}
F(r\cos\phi,r\sin\phi)&=\int\limits_{-\infty}^{\infty}\bigg[\int\limits_{-\infty}^\infty \int\limits_{-\infty}^\infty f(x,y)\delta(\rho-x\cos\phi-y\sin\phi)dxdy\bigg]e^{-2\mi\pi\rho r}d\rho\\
&=\int\limits_{-\infty}^\infty\breve{f}(\rho,\phi)e^{-2\mi\pi\rho r}d\rho.
\end{split}
\end{align}
\end{widetext}
Thus, a one-dimensional Fourier transform of the FT gives the spectrum, subsequently giving $f(x,y)$. In our case, the one-dimensional transformation of the parallel projections defined by the Radon transform $\breve{f}(\rho,\phi)$ makes possible  the reconstruction of the WDF. Consequently, for the discrete case, we have that the DFT of the parallel projection defined by the distance $Z$ is
\begin{equation}
F_n(Z_m)=\sum_{k=0}^{2N}I_k(Z_m)e^{-\mi\frac{2\pi}{N}nk},
\end{equation} 
with $n\in\mathbb{N}$, and $Z_m$ is the $m$-th distance in a range of $Z\in[Z_0,Z_1,\dots,Z_m,\dots]$. Therefore, a point in phase space $(x_i,y_i)$ is determined by the discrete 2D Fourier transform, such that
\begin{align}
\begin{split}
W_g(x_i,y_i)&=\frac{1}{\sqrt{(2N+1)(2M+1)}}\\
&\times\sum_{n=0}^{2N}\sum_{m=0}^{2M}F_n(Z_m)e^{2\mi\pi(\frac{nx_i}{2N}+\frac{my_i}{2M})}.
\end{split}
\end{align}
This relationship summarizes the sketch shown in Fig.~\ref{Radon_fig}. Once we obtain the 1D Fourier transform of the projection of any function at a given distance (angle), $F_{n}(Z_m)$, we can obtain the Wigner function at any point in the phase space following the same algorithm. Another famous inversion scheme is the Filtered Back projection \cite{Bruyant01102002}. It is derived from Eq.~\eqref{C2}.\\

\begin{acknowledgments} 
The authors acknowledge the valuable guidance of Armando P\'erez-Leija in the early stages of this work. A.R.U. acknowledges support from DGAPA-UNAM posdoctoral program (POSDOC) 2023-2024, and to ICF-UNAM for the in-place support.\\

$^\dagger$ This paper is dedicated to the memory of our colleague Dr. Gustavo Rodr\'{i}guez-Zurita.
\end{acknowledgments}

\bibliography{Wigner-Radon}
\end{document}